\documentclass[onecolumn,showpacs,showkeys,amsmath,amssymb]{revtex4}

\usepackage{graphics}
\usepackage{graphicx}
\usepackage{epsfig}
\usepackage{mathrsfs,bbm}

\newcommand{\ot}[0]{\otimes}
\newcommand{\ket}[1]{|#1\rangle}
\newcommand{\kket}[1]{\left|#1\right>}

\newcommand{\bra}[1]{\langle#1|}

\newcommand{\proj}[1]{\ket{#1}\bra{#1}}
\newcommand{\av}[2]{\bra{#2}{#1}\ket{#2}}

\newcommand{\Tr}[0]{\mathrm{Tr}}

\begin{document}
\title{Rotationally invariant bipartite states and bound entanglement}
\author{Remigiusz Augusiak}
\author{Julia Stasi\'nska}
\email{jul_sta@wp.pl}
 \affiliation{Faculty of Applied Physics
and Mathematics, Gda\'nsk University of Technology, Gda\'nsk,
Poland}
\date{\today}
\begin{abstract}
We consider rotationally invariant states in
$\mathbb{C}^{N_{1}}\ot \mathbb{C}^{N_{2}}$ Hilbert space with even
$N_{1}\geq 4$ and arbitrary $N_{2}\geq N_{1}$, and show that in
such case there always exist states which are inseparable and
remain positive after partial transposition, and thus the PPT
criterion does not suffice to prove separability of such systems.
We demonstrate it applying a map developed recently by Breuer
[H.-P. Breuer, Phys. Rev. Lett {\bf 97}, 080501 (2006)] to states
that remain invariant after partial time reversal.
\end{abstract}
\maketitle
\section{Introduction}
\label{Sec.I}
One of the most important problems of rapidly developing branch of
science Quantum Information Theory \cite{NCh} is to determine
whether a given quantum state is separable or entangled. We say
that a given state $\varrho$ acting on a finite dimensional
product Hilbert space $\mathcal{H}_{1}\ot\mathcal{H}_{2}$ is
separable or classically correlated if it can be written as a
convex linear combination of product density operators
\cite{Werner}, i.e.,
\begin{equation}
\varrho=\sum_{n}p_n \varrho_{n}^{(1)} \otimes  \varrho_{n}^{(2)},
\end{equation}
where $\varrho_{n}^{(1(2))}\in\mathcal{B}(\mathcal{H}_{1(2)})$ for
all $n$ and $p_n$ are nonnegative coefficients fulfilling the
condition $\sum_{n}p_n=1$. Otherwise the state is called entangled
or inseparable. An important necessary and sufficient criterion
used to check the separability of a state was developed by the
Horodecki's \cite{PH1}. It applies the positive but not completely
positive maps, precisely, it states that a density matrix
$\varrho$ is separable if and only if the operator $(I\otimes
\Lambda)(\varrho)$ is positive for all positive but not completely
positive maps $\Lambda: \mathcal{B}(\mathcal{H}_{2})\rightarrow
\mathcal{B}(\mathcal{H}_{1})$. Even though making use of the
criterion is not an easy task, together with the result of Peres
\cite{Peres} it provides a strong condition for separability,
based on partial transposition. It says that $\varrho$ is
entangled if it has a nonpositive partial transposition. For low
dimensional systems such as $2\ot 2$ and $2\ot 3$ this is also a
sufficient criterion for separability \cite{PH1}, however, in
general there exist states that have positive partial
transposition (PPT) and simultaneously are entangled \cite{BE1}.
Such operators are certainly nondistillable \cite{BE2} and belong
to the class of bound entangled (BE) states.

An interesting question is thus whether there exist classes of
states (except for $2\ot2$ and $2\ot 3$ systems) where partial
transposition provides a necessary and sufficient test for
separability. One could expect that invariance of states under
certain group of symmetry would lead to some interesting results.
Such states have a relatively simple structure and therefore have
been studied extensively in the literature
\cite{Werner1,Sym2,Sym1,diplomarbeit,Schliemann1,Schliemann2,4x4,3xN,Sym5,Sym3,Sym4}.
In particular, it was shown recently that positive partial
transposition is a sufficient criterion in the case of $2\otimes
N$ rotationally invariant states \cite{Schliemann1,Schliemann2}
and $3\otimes N$ rotationally invariant states with integer total
angular momentum \cite{4x4,3xN}. On the other hand in Ref.
\cite{B-criterion} it was shown that for $N\ot N$ rotationally
invariant systems with even $N\geq 4$ there always exist bound
entangled states detected by certain map.

It is the purpose of the present paper to consider the
separability of more general $SO(3)$ invariant states. We
concentrate on $N_{1}\ot N_{2}$ systems with even $N_{1} \geq 4$
and arbitrary $N_{2}\geq N_{1}$ and show that there is always a
region in the PPT set where states are bound entangled proving,
thus, that in such systems positive partial transposition is only
a necessary condition for separability. We achieve this using a
recently introduced positive indecomposable map $\Phi$ given by
Eq.(\ref{mapphi}) \cite{B-criterion}. The map belongs to the class
of indecomposable positive maps arising from the reduction
criterion \cite{reduction,red2} which was studied in detail by
Hall \cite{Hall}. The action of $\Phi$ on a given operator $B$
from $\mathcal{B}(\mathbb{C}^{N})$ is such that
\begin{equation}\label{mapphi}
\Phi (B)=(\Tr{B})\mathbbm{1}_{N}-B-\vartheta (B),
\end{equation}
where $\vartheta$ denotes the time reversal operation and
$\mathbbm{1}_{N}$ is a $N\times N$ identity matrix. The map $\Phi$
is positive if $N$ is an even number and therefore leads to the
following necessary condition for separability on
$\mathbb{C}^{N_{1}}\ot\mathbb{C}^{N_{2}}$ Hilbert space with even
$N_{1}$:
\begin{equation}\label{B-op}
\Phi_{1} (\varrho) \equiv (\Phi \otimes I)(\varrho) \geq 0.
\end{equation}

We show that when applying the Breuer's map to rotationally
invariant states one can restrict to a family of operators
invariant under partial time reversal. Moreover the results
obtained for $\vartheta_1$-invariant states can be extended to the
PPT rotationally invariant states. We also show that in our case
the set of states invariant under partial time reversal can be
easily found and therefore it is possible to prove the existence
of BE states in higher dimensional rotationally invariant systems.

The paper is organized as follows. In Sec. \ref{Sec.II} we give a
brief description of representations of $SO(3)$-invariant states
and action of certain positive but not completely positive maps on
this class of states. In Sec. \ref{Sec.III} we present a special
case of $4\ot N$ and on the basis of this example analyze more
general case $N_1\ot N_2$ with even $N_1 \geq 4$ and arbitrary
$N_{2}\geq N_{1}$. In particular we show that to determine the set
of BE states detected by the Breuer's map one can restrict to the
subset of states invariant under partial time reversal and then
extend the result to the set of PPT states. The results also
suggest that the map (\ref{mapphi}) could provide both necessary
and sufficient separability criterion for states invariant under
partial time reversal.

\section{Rotationally invariant states}
\label{Sec.II}
\subsection{Representations}
Assume that we are given a bipartite quantum state represented by
a density matrix $\varrho$ acting on a finite product Hilbert
space
$\mathcal{H}_{1}\ot\mathcal{H}_{2}=\mathbb{C}^{N_{1}}\ot\mathbb{C}^{N_{2}}$
such that $N_1\leq N_2$. The angular momenta of the particles are
$j_{1}=(N_{1}-1)/2$ and $j_{2}=(N_{2}-1)/2$, respectively. The
characterization of entanglement of such states is in general a
difficult task. However, it simplifies when one impose some
constraints on considered density matrices. Hereafter we shall be
assuming that $\varrho$ is invariant under the action of $SO(3)$
group. More rigorously this assumption means that the following
relation holds
%
\begin{equation}\label{inv}
\left[\mathcal{D}^{(j_{1})}(R)\ot\mathcal{D}^{(j_{2})}(R),\varrho\right]=0
\end{equation}
for all proper rotations $R$ from $SO(3)$ group. Here
$\mathcal{D}^{(j_{1(2)})}(R)$ denote the unitary irreducible
representation of the $SO(3)$ group on the respective state
spaces. In view of the Shur's lemma, the states which obey the
above condition, can be written in the following form
\begin{equation}\label{alpha}
\varrho=\frac{1}{\sqrt{N_{1}N_{2}}}\sum_{J=|j_{1}-j_{2}|}^{j_{1}+j_{2}}\frac{\alpha_{J}}{\sqrt{2J+1}}\,P_{J},\quad
P_J=\sum_{M=-J}^{J}\proj{JM},
\end{equation}
where $\ket{JM}$ are the common eigenvectors of the square of
total angular momentum operator and of its $z$-component. Thus the
set of rotationally invariant states is isomorphic to a proper
subset of vectors $\alpha$ from $\mathbb{R}^{N_{1}}$, which are
nonnegative, i.e., $\alpha_{J}\geq 0$ and fulfil the following
normalization condition
\begin{equation}
\sum_{J=|j_{1}-j_{2}|}^{j_{1}+j_{2}}\sqrt{\frac{2J+1}{N_{1}N_{2}}}\,\alpha_{J}=1.
\end{equation}
On the other hand, as proposed by Breuer in Refs. \cite{4x4,3xN},
each rotationally invariant state can be written as a combination
of Hermitian operators
\begin{equation}\label{qk}
Q_K=\sum_{q=-K}^{K}T_{K,q}^{(1)}\otimes T_{K,q}^{(2)\dagger},
\qquad K=0,1,\ldots,2j_1,
\end{equation}
where $T_{K,q}^{(i)}$ are the components of an irreducible tensor
operator \cite{Edmonds}. Subscript $K$ is the rank of this tensor
operator and $q$ takes on $2K+1$ values, $q=-K,\ldots,K$. The
operators $Q_{K}$ are rotationally invariant and form a complete
set in the space of rotationally invariant states. Thus any
$SO(3)$ invariant state can be written as their linear
combination, i.e.,
\begin{equation}\label{beta}
\varrho=\frac{1}{\sqrt{N_{1}N_{2}}}\sum_{K=0}^{2j_{1}}\frac{\beta_{K}}{\sqrt{2K+1}}\,Q_{K}.
\end{equation}
Like in the case of {$P_{J}$} representation the state is uniquely
characterized by a parameter vector $\beta \in
\mathbb{R}^{N_{1}}$. Due to the fact that tensor operators $Q_K$
are traceless for $K\neq 0$ the normalization condition gives
$\beta_0=1$. As it was shown in Ref. \cite{3xN}, the
$N_1$-dimensional vectors $\alpha$ and $\beta$ are related by a
linear transformation $\beta=L \alpha$ with matrix elements of the
orthogonal matrix L given by
\begin{equation}\label{lmatrix}
L_{KJ}=\sqrt{(2K+1)(2J+1)}(-1)^{j_1+j_2+J}\left\{\begin{array}{ccc}
  j_1 & j_2 & J \\
  j_2 & j_1 & K \\
\end{array}\right\}.
\end{equation}

The reason for introducing two different representations is their
convenience for certain purposes. $P_J$-representation is suitable
to determine the state space, whereas the operation of partial
time reversal, unitarily equivalent to the partial transposition,
can be performed much more easily in the ${Q_K}$ basis. Thus we
apply the latter to identify the set of rotationally invariant
states with positive partial time reversal which is the same as
set of states that remain positive after partial transposition.
\subsection{Separability}
To determine the set of rotationally invariant PPT states (from
now on denoted by $\mathcal{R}_{\mathrm{ppt}}$) we apply a map
unitarily equivalent to partial transposition, i.e., the partial
time reversal map \cite{3xN}. It is more useful in our case due to
the fact that, unlike the partial transposition, it preserves the
$SO(3)$ invariance of the state. The operation of partial time
reversal acts as follows
\begin{equation}
\vartheta_1(B)=(\vartheta \ot I)(B),
\end{equation}
where $\vartheta$ is the time reversal map acting on a given
operator $B$ as $\vartheta(B)=
 V\,B^{T}\,V^{\dagger}$.  Here $V$ is a unitary rotation by the
angle $\pi$ about the $y$-axis. As mentioned before, the action of
partial time reversal operator is especially simple in the $Q_K$
representation
\begin{equation}\label{part}
\vartheta_{1}:\beta_{K}\to (-1)^{K} \beta_{K},
\end{equation}
which follows directly from the relation $\vartheta_1
(Q_{K})=(-1)^{K} Q_{K}$ and equation (\ref{beta}). Let us recall
one more property of $\vartheta_{1}$ map, i.e., preservation of
separability. It means that if $\varrho$ is separable then
$\vartheta_{1}(\varrho)$ is also separable. We apply this fact in
the next section to prove the separability of certain PPT states.
%

Since the set of separable states is a subset of PPT states, one
should have the condition unambiguously determining the
separability of analyzed states. A method useful to identify the
separable invariant states was developed in \cite{Werner1} and is
based on the action of a projection super-operator $\Pi$. The
super-operator for $SO(3)$ group of symmetry, considered in this
paper, projects each state onto a rotationally invariant state
space as follows
\begin{equation}\label{pi}
\Pi (\varrho)=\sum_{J=|j_1-j_2|}^{j_1+j_2} \frac{\Tr (P_J
\varrho)}{2J+1} P_J=\sum_{K=0}^{2 j_1} \frac{\Tr (Q_K
\varrho)}{2K+1} Q_K,
\end{equation}
preserving the separability of a state. The last property allows
us to identify the set of separable states
$\mathcal{R}_{\mathrm{sep}}$ with a convex linear combination of
$\Pi$-projections of pure normalized product states
$\ket{\phi^{(1)}\phi^{(2)}}\in
\mathbb{C}^{N_{1}}\ot\mathbb{C}^{N_{2}}$:
\begin{equation}
\mathcal{R}_{\mathrm{sep}}=\mathrm{conv}\left\{\varrho:\varrho=\Pi\left(
P_{\phi^{(1)}}\ot P_{\phi^{(2)}}\right)
\right\},
\end{equation}
where $P_{\phi^{(i)}}$ is the projection onto the state
$\ket{\phi^{(i)}}$. Consequently to find separable states among
PPT states it is enough to show that for the extreme points of
$\mathcal{R}_{\mathrm{sep}}$ (denoted by $\varrho_{\mathrm{ext}}$)
there exist pure normalized product states satisfying relation
$\Pi( P_{\phi^{(1)}}\ot P_{\phi^{(2)}})=\varrho_{\mathrm{ext}}$.
However to find the ''suspected'' extreme points
$\varrho_{\mathrm{ext}}$ one should have a criterion detecting at
least some bound entanglement, so that one could look for
separable states in a smaller set.

In our case the map developed recently in \cite{B-criterion}
proved to be especially useful. It leads to the necessary
criterion (\ref{B-op}). Using this criterion we show that the set
of PPT rotationally invariant states for even $N_1 \leq N_2$
always contains entangled states. To prove this it is enough to
check the criterion on the subset of $\vartheta_{1}$-invariant
states
\begin{equation}\label{t-inv}
\varrho_{\mathrm{inv}}=\frac{1}{2}[\varrho+\vartheta_{1}(\varrho)],
\quad \varrho \in \mathcal{R}_{\mathrm{ppt}},
\end{equation}
since the action of Breuer's map is the same for all states
$\varrho$ satisfying above equation and equivalent to
\begin{equation}\label{inv2}
\Phi_1(\varrho)=\mathbbm{1}_{N_1}\ot
\Tr_1(\varrho)-\varrho-\vartheta
_{1}(\varrho)=\frac{1}{N_{2}}\mathbbm{1}_{N_1}\ot
\mathbbm{1}_{N_2}-2 \varrho_{\mathrm{inv}}.
\end{equation}
In the second equality appearing in Eq. (\ref{inv2}) we make use
of the fact that $\Tr_1(\varrho)=(1/N_{2})\mathbbm{1}_{N_{2}}$
for all rotationally invariant states. As a result of the above
relations we can simplify the criterion (\ref{B-op}) to
\begin{equation}\label{B-opinv}
\frac{1}{N_{2}}\mathbbm{1}_{N_1}\ot \mathbbm{1}_{N_2}-2
\varrho_{\mathrm{inv}}\geq 0.
\end{equation}
It automatically follows that if $\varrho_{\mathrm{inv}}$ fulfils
the criterion (\ref{B-opinv}) then all $\varrho$'s from equation
(\ref{t-inv}) satisfy (\ref{B-op}).

\section{Rotationally invariant states and bound entanglement}
\label{Sec.III}
In this section we shall consider rotationally invariant states in
the context of bound entanglement. We show that in
$\mathbb{C}^{N_{1}}\ot\mathbb{C}^{N_{2}}$ Hilbert space with even
$N_1$ and arbitrary $N_2\geq N_1$ there always exist PPT entangled
states.

At the very beginning to provide some insight into the state space
structure we focus on the case of the $4\ot N$ system, since it
may be easily visualized in $\mathbb{R}^3$.
\subsection{$\mathbf{4\ot N}$ system}
Let us now restrict our attention to the case of $N_{1}=4$
($j_1=3/2$) and arbitrary $N_{2}=2j_{2}+1\geq N_{1}$ from now on
denoted by $N$. In this paragraph we will also denote $j_2$ by
$j$. The total angular momentum of the system $J$ takes on the
values $J=j-3/2,\ldots,j+3/2$ and thus every rotationally
invariant state is represented by 4 coordinates which satisfy the
conditions $\alpha_{J}\geq 0$ and
\begin{equation}
\sum_{J=j-3/2}^{j+3/2}\sqrt{\frac{2J+1}{4N}}\alpha_{J}=1.
\end{equation}
By the normalization condition the number of independent
parameters $\alpha_J$ describing a state can be reduced to 3.
\noindent In order to characterize this set of rotationally
invariant states we need to find its extreme points in $\alpha$
space. Straightforward calculations lead to
\begin{equation*}\label{alfapoints}
A_{\alpha}=\left(\sqrt{\frac{4N}{N-3}},0,0,0\right),\quad
B_{\alpha}=\left(0,\sqrt{\frac{4N}{N-1}},0,0\right),\quad
C_{\alpha}=\left(0,0,\sqrt{\frac{4N}{N+1}},0\right),\quad
D_{\alpha}=\left(0,0,0,\sqrt{\frac{4N}{N+3}}\right).
\end{equation*}
To determine the set of PPT states we transform the vectors
$\mathbf{\alpha}$ to $\mathbf{\beta}$ using the linear
transformation $L$ which matrix elements are given by Eq.
(\ref{lmatrix}). In the considered case of $4\ot N$ the
transformation matrix is of the form
\begin{equation}
\hspace{-1cm}L=\frac{1}{2}\left(
\begin{array}{cccc}
\sqrt{\frac{N-3}{N}} & \sqrt{\frac{N-1}{N}} & \sqrt{\frac{N+1}{N}}
& \sqrt{\frac{N+3}{N}} \\
-3\sqrt{\frac{(N-3)(N+1)}{5(N-1)N}} & -\frac{N+7}{\sqrt{5N(N+1)}}
& \frac{N+7}{\sqrt{5N(N-1)}} &
-3\sqrt{\frac{(N+3)(N-1)}{5(N+1)N}}\\
\sqrt{\frac{(N-3)(N+1)(N+2)}{N(N-1)(N-2)}} &
\frac{(N-5)\sqrt{N+2}}{\sqrt{(N-2)N(N+1)}} &
-\frac{(N+5)\sqrt{N-2}}{\sqrt{(N-1)N(N+2)}} &
\sqrt{\frac{(N-1)(N-2)(N+3)}{N(N+1)(N+2)}} \\
-\sqrt{\frac{(N+1)(N+2)(N+3)}{5(N-2)(N-1)N}} &
3\sqrt{\frac{(N^{2}-9)(N+2)}{(N-2)N(N+1)}} &
-3\sqrt{\frac{(N^{2}-9)(N-2)}{(N-1)N(N+2)}} &
\sqrt{\frac{(N-3)(N-2)(N-1)}{5N(N+1)(N+2)}}
\end{array}
\right).
\end{equation}

At the same time we observe, that the 4-dimensional vectors
$\beta$ are unambiguously characterized by three coordinates since
$\beta_{0}=1$ for all rotationally invariant density matrices.
This allows us to restrict our considerations to the three
parameters $\beta_1,\beta_2,\beta_3$ and visualize all considered
sets in $\mathbb{R}^{3}$. The $L$ transformation carried on the
extreme points gives us the vertices of the tetrahedron in the
space of parameters $\beta_1,\beta_2,\beta_3$ as follows
\begin{eqnarray*}
A&=&\left(-3\sqrt{\frac{N+1}{N-1}},\sqrt{\frac{(N+1)(N+2)}{(N-1)(N-2)}},\sqrt{\frac{(N+1)(N+2)(N+3)}{5(N-1)(N-2)(N-3)}}\right),\nonumber\\
B&=&\left(-\frac{N+7}{N+1}\sqrt{\frac{N+1}{5(N-1)}},-\frac{N-5}{N+1}\sqrt{\frac{(N+1)(N+2)}{(N-1)(N-2)}},3\sqrt{\frac{(N+2)(N+3)(N-3)}{5(N-2)(N-1)(N+1)}}\right),\nonumber\\
\end{eqnarray*}
\begin{eqnarray*}
C&=&\left(\frac{N-7}{N+1}\sqrt{\frac{N+1}{5(N-1)}},-\frac{N+5}{N-1}\sqrt{\frac{(N-1)(N-2)}{(N+1)(N+2)}},-3\sqrt{\frac{(N-2)(N-3)(N+3)}{5(N+2)(N-1)(N+1)}}\right),\nonumber\\
D&=&\left(3\sqrt{\frac{N-1}{5(N+1)}},\sqrt{\frac{(N-2)(N-1)}{(N+2)(N+1)}},\sqrt{\frac{(N-1)(N-2)(N-3)}{5(N+1)(N+2)(N+3)}}\right).
\end{eqnarray*}
\begin{figure}[!hbp]
\begin{center}
  \includegraphics[width=0.7\textwidth]{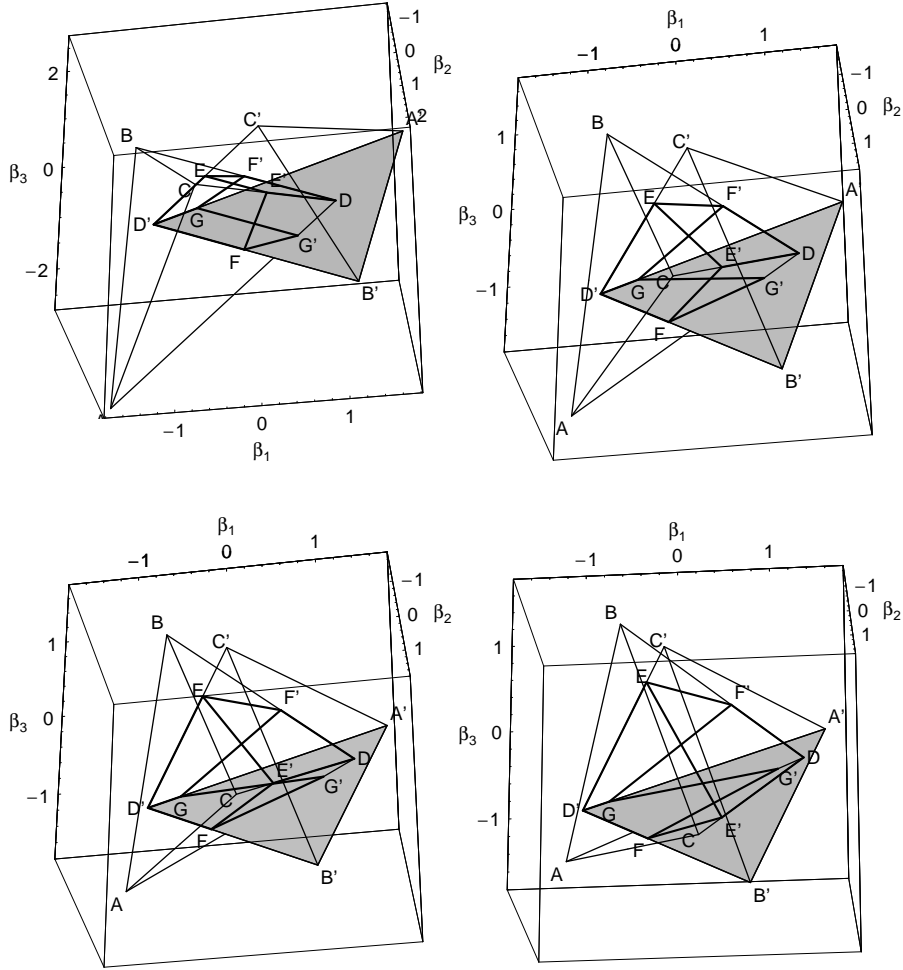}
\end{center}
  \caption{The set of rotationally invariant states in $\mathbb{C}^{4}\ot\mathbb{C}^{N}$ (tetrahedron
  $ABCD$), its image under partial time reversal $(A'B'C'D')$, and
  intersection of the sets $(DD'EE'FF'GG')$ for various values of
  $j$, namely $j=3/2$ (left upper), $j=5/2$ (right upper), $j=7/2$ (left lower), $j=11/2$ (right lower).
  The PPT region becomes larger with the growth of dimension of the second subsystem and the PPT criterion fails to
  detect many entangled states with the growth of asymmetry between the dimensions of subsystems.}\label{intersection}
\end{figure}

Now we can find the image of the tetrahedron $ABCD$ under the
action of $\vartheta_1$. The extreme points of this set transform
according to the relation (\ref{part}), consequently only the sign
of $\beta_1$ and $\beta_3$ coordinates change. We denote the
points corresponding to $A,\,B,\,C,\,D$ by $A',\,B',\,C'$ and
$D'$, respectively. To identify the set of PPT states we need to
find the intersection of the tetrahedrons $ABCD$ and $A'B'C'D'$.
Straightforward calculations lead to the set $DD'EE'FF'GG'$ (see
Fig. \ref{intersection}). One may easily verify that the pairs of
points $(E,E'),(F,F'),(G,G')$ are the images of each other under
$\vartheta_{1}$ (e.g. $E'=\vartheta_{1}(E)$) and thus it is
sufficient to find coordinates of the points $E$, $F$, and $G$,
which are as follows

\begin{eqnarray*}
E&=&\left(-\sqrt{\frac{N-1}{5(N+1)}},-\sqrt{\frac{(N-1)(N-2)}{(N+1)(N+2)}},3\sqrt{\frac{(N-1)(N-2)(N-3)}{5(N+1)(N+2)(N+3)}}\right),\nonumber\\
F&=&\left(-\frac{9(N-4)}{7N-20}\sqrt{\frac{N-1}{5(N+1)}},\frac{N+4}{7N-20}\sqrt{\frac{(N-1)(N-2)}{(N+1)(N+2)}},-\frac{13N+28}{7N-20}\sqrt{\frac{(N-1)(N-2)(N-3)}{5(N+1)(N+2)(N+3)}}\right),\nonumber\\
G&=&\left(-\frac{3(N+1)}{N+5}\sqrt{\frac{N-1}{5(N+1)}},\frac{N+2}{N+5}\sqrt{\frac{(N-1)(N+1)}{(N-2)(N+2)}},-\frac{N-7}{N+5}\sqrt{\frac{(N-1)(N+1)(N+2)}{5(N-3)(N+3)(N-2)}}\right).
\end{eqnarray*}
The described sets are presented in Fig. \ref{intersection} for
$N=4,\,6,\,8,\,12$. One could see that the overlap of tetrahedrons
$ABCD$ and $A'B'C'D'$ grows with the increase of $N$. This
immediately leads to the conclusion that the NPT set shrinks with
increasing $N$. Now applying the methods described in the previous
section we characterize the separability of PPT states.

\begin{figure*}[!hbp]
\begin{center}
  \includegraphics[width=0.7\textwidth]{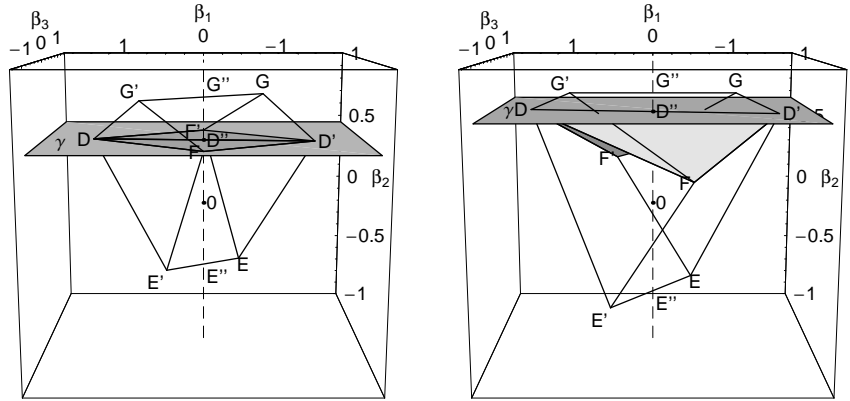}\\
\end{center}
  \caption{The set of rotationally invariant PPT states for $j=3/2$ and $3$, respectively.
  The $\gamma$ plain labelled in the pictures is the boundary of the region in which entanglement is detected by
  Breuer's map. BE states which can be detected by the Breuer's map lie above the $\gamma$ plain in the set of PPT operators.}\label{PPTset}
\end{figure*}

Firstly we apply the Breuer's criterion to the subset of
$\vartheta_1$-invariant states, namely states represented by
points lying on the line $E''G''$ (see Fig. \ref{PPTset})
\begin{equation}
\varrho_{\mathrm{inv}}(t)=(1-t)E''+tG'', \qquad t\in [0,1].
\end{equation}
The criterion (\ref{inv2}) applied to $\varrho_{\mathrm{inv}}(t)$
leads to the operator with $\alpha_{j-3/2}$ given by:
\begin{equation}\label{amin}
\alpha_{j-3/2}(t)=\sqrt{\frac{N-3}{N}}\left(1-\frac{(N-1)(N+4)}{(N-2)(N+5)}t\right).
\end{equation}
As it may be easily verified that remaining $\alpha_J$'s are
nonnegative for all values of $t$ and thus the nonpositivity
condition reduces to
\begin{equation}\label{cond}
1-\frac{(N-1)(N+4)}{(N-2)(N+5)}t<0.
\end{equation}
States $\varrho(t)$ for $t$ satisfying the inequality are
entangled. The parameter $t$ for which the LHS of (\ref{cond})
equals zero represents the point $D''$ lying in the middle of the
line $DD'$. At the same time the inequality is satisfied by $t=1$
which implies that points between $D''$ and $G''$, including $G''$
represent entangled states.
Now we denote by $\gamma$ the plain perpendicular to the line
$E''G''$ and intersecting point $D''$ (Fig. \ref{PPTset}). From
relation (\ref{B-opinv}) and a remark below it we can immediately
conclude that all points lying in the PPT set above the $\gamma$
plain are entangled.

Now we move to the points for which $\Phi_1(\varrho)\geq 0$ (such
points lie on and below $\gamma$) and use the $\Pi$ projection
argument introduced in Sec II.B to show the separability of $D$
and $E$ (the separability of $D',\,E'$ results immediately from
the properties of $\vartheta_{1}$ map, namely preservation of
separability).

Let us first recall the special symmetric case $4 \otimes 4$
solved in \cite{4x4,diplomarbeit}. In this case points
$D,\,D',\,F,\,F'$ lie on the $\gamma$ plain and are all separable.
The points $E,\,E'$ are also separable and thus the $\gamma$ plain
is a boundary between the separable and bound entangled region.
\begin{figure*}[!hbp]
\begin{center}
  \includegraphics[width=0.4\textwidth]{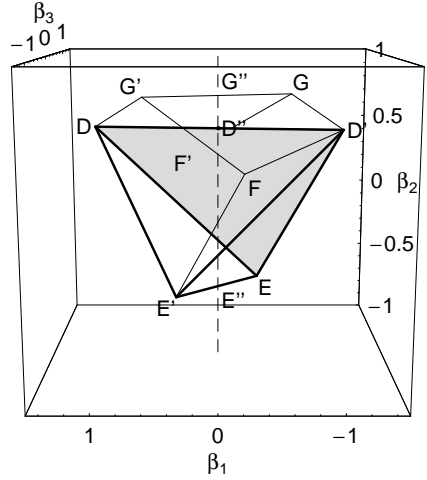}\\
\end{center}
  \caption{The set of rotationally invariant PPT states for $j=2$. The borders of the minimal
  separable region $DD'EE'$ are marked with thick lines.}\label{minimal}
\end{figure*}

In the general case $4\otimes N$, the full characterization of the
separable set is more difficult. Following the method developed by
Breuer \cite{3xN} we show that certain separable pure product
states $\ket{\phi^{(1)}\phi^{(2)}}$ are projected by $\Pi$ (see
(\ref{pi})) onto points $D$ and $E$. For this purpose we take the
functionals
\begin{equation}\label{bfun}
\tilde{\beta}_K[\phi^{(1)},\phi^{(2)}] =\sqrt{\frac{4N}{2K+1}}
\sum_{q=-K}^{K}\av{T_{K,q}^{(1)}}{\phi^{(1)}}\av{T_{K,q}^{(2)\dagger}}{\phi^{(2)}},
\end{equation}
which, applying relation (\ref{pi}), map pure product states into
the $\beta$ space. Namely the $\tilde{\beta}_K$ functional is the
$\beta_K$ coordinate of a pure product state
$\ket{\phi^{(1)}\phi^{(2)}}$ after action of $\Pi$ projection. The
matrix elements of $T_{K,q}$ used in the above equation are
related to Wigner $3$-$j$ symbol by the formula \cite{Edmonds}:
\begin{equation}
\bra{j,m}T_{K,q}\ket{j,m'}=\sqrt{2K+1}(-1)^{j-m}\left(\begin{array}{ccc}
  j & j & K \\
  m & m' & -q \\
\end{array}\right)
\end{equation}
and by the relation $T_{K,q}^{\dagger}=(-1)^q T_{K,-q}$ one can
determine the matrix elements of the conjugate.

To prove that $E$ is separable it suffices to consider the states
\begin{equation}
\ket{\tilde{\phi}^{(1)}}=\kket{\frac{3}{2},-\frac{1}{2}},\quad
\ket{\tilde{\phi}^{(2)}}=\kket{\frac{N-1}{2},\frac{N-1}{2}}.
\end{equation}
Due to the selection rules for $3$-$j$ symbol the only
nonvanishing elements of the sum in (\ref{bfun}) are those with
$q=0$, hence
\begin{eqnarray*}
\tilde{\beta}_{1}[\tilde{\phi}^{(1)},\tilde{\phi}^{(2)}]&=&\sqrt{\frac{4N}{3}}\av{T_{1,0}^{(1)}}{\tilde{\phi}^{(1)}}\av{T_{1,0}^{(2)\dagger}}{\tilde{\phi}^{(2)}}=-\sqrt{\frac{N-1}{5(N+1)}},\nonumber\\
\tilde{\beta}_{2}[\tilde{\phi}^{(1)},\tilde{\phi}^{(2)}]&=&\sqrt{\frac{4N}{5}}\av{T_{2,0}^{(1)}}{\tilde{\phi}^{(1)}}\av{T_{2,0}^{(2)\dagger}}{\tilde{\phi}^{(2)}}=-\sqrt{\frac{(N-1)(N-2)}{(N+1)(N+2)}},\nonumber\\
\tilde{\beta}_{3}[\tilde{\phi}^{(1)},\tilde{\phi}^{(2)}]&=&\sqrt{\frac{4N}{7}}\av{T_{3,0}^{(1)}}{\tilde{\phi}^{(1)}}\av{T_{3,0}^{(2)\dagger}}{\tilde{\phi}^{(2)}}=3\sqrt{\frac{(N-1)(N-2)(N-3)}{5(N+1)(N+2)(N+3)}}.\nonumber\\
\end{eqnarray*}
These are indeed the coordinates of $E$ which implies that $E$
represents the separable state.

To prove the separability of $D$ we follow the arguments given by
Breuer in \cite{3xN}. Since $D$ is an extreme point of the set of
$SO(3)$ invariant states it has a single nonzero coordinate in
$\alpha$-space. The nonzero coordinate corresponds to the largest
$J=J_{\mathrm{max}}=j_1+j_2$. Consequently the spectral
decomposition of the state always contains the projection on
$\ket{J_{\mathrm{max}},J_{\mathrm{max}}}$ which is a separable
pure product state $\ket{j_1,j_1}\ot\ket{j_2,j_2}$. This implies
that for arbitrary $j_1,j_2$ one can find a separable state
$\ket{j_1,j_1}\ot\ket{j_2,j_2}$ which is mapped under $\Pi$ to the
state represented by point D.

The tetrahedron with vertices $DD'EE'$ in Fig. (\ref{minimal})
represents the minimal separable set (see also
\cite{diplomarbeit}). The separability of the points $F$ and $F'$
as well as points lying between $\gamma$ and thetetrahedron
$DD'EE'$ within the PPT set is still undetermined. However in the
case of $4\ot 5$ considered by Hendriks in \cite{diplomarbeit} the
set of separable states is given not only by points $DD'EE'$ but
extends to the region near points $F$ and $F'$. The separable
states found numerically by Hendriks lie on lines $E'F$ and $EF'$.
This result allows us to suspect that for $N \geq 5$ there also
exist separable states outside the tetrahedron $DD'EE'$.

\subsection{Bound entanglement in higher dimensional systems}
Let us now move on to the general case of
$\mathbb{C}^{N_{1}}\ot\mathbb{C}^{N_{2}}$ Hilbert space with even
$N_1 \geq 4$ and arbitrary $N_2\geq N_1$. For the purpose of
proving the existence of BE in this state space we confine
ourselves to the subset of $\vartheta_1$-invariant operators.

A set of rotationally and $\vartheta_1$-invariant states
($\mathcal{R}_{\mathrm{inv}}$) can be easily determined in the
$\beta$-space because the action of $\vartheta_{1}$ in this
representation is just a change of sign for $\beta_{K}$ with odd
$K$. As a result the states which remain unchanged after the
partial time reversal must have coordinates indexed by odd $K$'s
equal zero. Moreover, a normalized state has always $\beta_{0}=1$,
so each rotationally and $\vartheta_1$-invariant state in
$\mathbb{C}^{N_{1}}\ot\mathbb{C}^{N_{2}}$ with even $N_1$ is fully
characterized by the set of $(N_1-2)/2$ parameters $\beta_{2K}$
($K=1,2,...,(N_{1}-2)/2$), i.e.,
\begin{equation}\label{thetainv}
\varrho_{\mathrm{inv}}=\left(1,0,\beta_2,0,\beta_4,\ldots,0,\beta_{N_1-2},0\right).
\end{equation}
To determine the range of the $\beta_{2K}$ parameters for which
vector $\beta$ represents a density operator we impose the
constraint of positivity. This can be easily done in $P_J$
representation. Firstly we employ the $N_{1}\times N_{1}$ matrix
$L^{-1}=L^{T}$ (\ref{lmatrix}) to find the $\alpha_J$ coordinates
of a $\vartheta_1$-invariant state. This gives
\begin{eqnarray}\label{hyper}
\alpha_{J}&=&
(-1)^{\frac{N_1+N_2-2}{2}+J}\sqrt{2J+1}\Big(\left\{\begin{array}{ccc}
  (N_1-1)/2 & (N_2-1)/2 & J \\
  (N_2-1)/2 & (N_1-1)/2 & 0 \\
\end{array}\right\}\nonumber\\
&+&\sum_{K=1}^{(N_1-2)/2}\sqrt{4K+1}\left\{\begin{array}{ccc}
  (N_1-1)/2 & (N_2-1)/2 & J \\
  (N_2-1)/2 & (N_1-1)/2 & 2K\\
\end{array}\right\}\beta_{2K}\Big).
\end{eqnarray}
The positivity condition $\alpha_J\geq 0$ leads to the set of
inequalities for $\beta_{2K}$ parameters. The $\beta_{2K}$
parameters describing the $SO(3),\,\vartheta_1$-invariant states
lie between the hyperplains given by equations $\alpha_{J}=0$,
where $\alpha_{J}$'s are given by (\ref{hyper}).

Our further reasoning follows from the structure of $4\otimes N$
space, where Breuer's map detects the entanglement of
$\vartheta_1$-invariant states above the $D''$ point.

The Breuer's map for rotationally, $\vartheta_1$-invariant states
has the form presented in Ineq. (\ref{B-opinv}). The identity
matrix in $\beta$ space is a vector with $\beta_0=N_1 N_2$ and
other coordinates equal zero. This follows from Eq. (\ref{beta})
and the fact that $Q_0$ operator is proportional to identity
($Q_0=(N_1 N_2)^{-1/2} \mathbbm{1}_{N_{1}}\otimes
\mathbbm{1}_{N_{2}}$). Therefore the $\beta$ vector obtained after
applying the $\Phi_1$ map and normalization is:
\begin{equation}
\frac{2}{2-N_{1}}\left(\frac{2-N_{1}}{2},0,\beta_2,0,\beta_4,\ldots,0,\beta_{N_1-2},0\right).
\end{equation}
One should notice that the vector is again $\vartheta_1$-invariant
and that the $\Phi_1$ map simply shifts the point with respect to
point $(1,0,\ldots,0)$.

We define the $\Gamma$ hyperplain:
\begin{equation}\label{Gamma}
\Gamma:\frac{1}{\sqrt{N_1 N_2}}+(-1)^{N_2}\frac{2}{N_1-2}
\sum_{K=1}^{(N_1-2)/2}\sqrt{4K+1} \left\{\begin{array}{ccc}
  (N_1-1)/2 & (N_2-1)/2 & (N_2-N_1)/2 \\
  (N_2-1)/2 & (N_1-1)/2 & 2K\\
\end{array}\right\}\beta_{2K}=0.
\end{equation}
It crosscuts the PPT $\vartheta_1$-invariant set in such way that
points lying on one side of it have always a negative eigenvalue
after the action of map (\ref{B-opinv}). To see this we apply the
$\Phi_1$ map to points lying on $\Gamma$. The obtained hyperplain
is the boundary of $\mathcal{R}_{\mathrm{inv}}$ given by
$\alpha_{(N_2-N_1)/2}=0$. So points lying on one side of $\Gamma$
are always mapped by $\Phi_1$ onto points outside the set of
positive operators, whereas points from the other side remain
positive.

Now we prove that $\Gamma$ always go through the interior of the
$\mathcal{R}_{\mathrm{inv}}$. To do it we find the analog of the
point $D''$ from the case $4\ot N$. We denote it by $\tilde{D}''$.
It corresponds to the state $[\varrho_{\mathrm{max}}+\vartheta_1(
\varrho_{\mathrm{max}})]/2$, where
$\varrho_{\mathrm{max}}=[N_1N_2(N_1+N_2-1)]^{-1/2}P_{(N_1+N_2-2)/2}$.
$\tilde{D}''$ is certainly $\vartheta_1$-invariant since it has
the form (\ref{t-inv}). It has all nonzero $\alpha_J$'s, which
implies that it belongs to the interior of
$\mathcal{R}_{\mathrm{inv}}$ (recall that the boundaries of the
set are given by $\alpha_J=0$). The even (and thus nonzero)
coordinates representing the $\tilde{D''}$ state in $\beta$ space
are:
\begin{equation}\label{de}
\beta_{2K}^{(\tilde{D}'')}=\sqrt{N_1
N_2(4K+1)}(-1)^{N_2}\left\{\begin{array}{ccc}
  (N_1-1)/2 & (N_2-1)/2 & (N_1+N_2-2)/2 \\
  (N_2-1)/2 & (N_1-1)/2 & 2K \\
\end{array}\right\},
\end{equation}
with $K=1,2,\ldots,(N_1-2)/2$. The coordinates fulfill the
equality (\ref{Gamma}) (the proof is given in appendix \ref{AppA})
so $\tilde{D}''$ always lies on the $\Gamma$ plain. This implies
that the $\Gamma$ hyperplain cuts through the set of PPT
$\vartheta_1$-invariant states. The above arguments prove the
existence of BE states for all $\vartheta_1$-invariant states from
$\mathbb{C}^{N_{1}}\ot\mathbb{C}^{N_{2}}$ Hilbert space with even
$N_1$ and arbitrary $N_2\geq N_1$. Moreover, by the relation
(\ref{t-inv}) our arguments immediately extend to the set of PPT
states.

\begin{figure*}[!hbp]
\begin{center}
  \includegraphics[width=\textwidth]{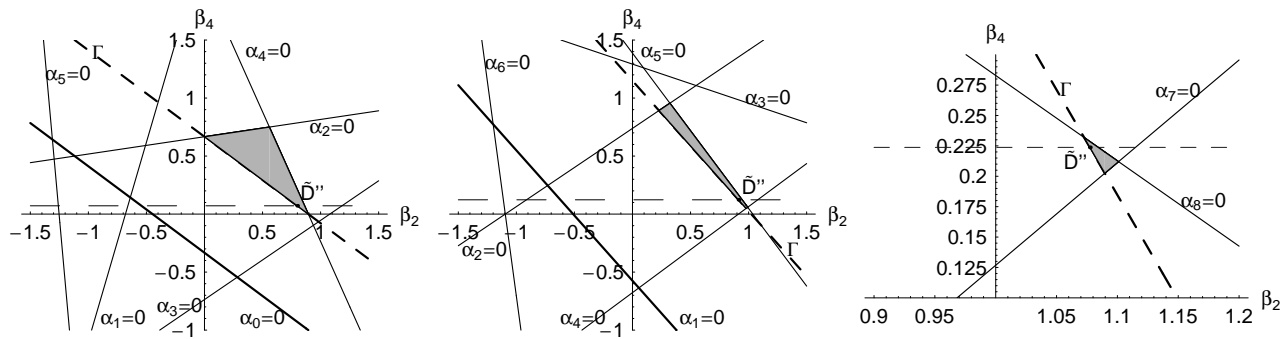}\\
\end{center}
  \caption{The subset of rotationally and $\vartheta_1$-invariant states for dimensions
  $6\ot 6$, $6\ot 8$ and $6\ot 14$, respectively. The constant lines are the borders
  of the $\vartheta_1$-invariant set, namely $\alpha_J=0$ $\left(J=(N_2-6)/2,...,(N_2+4)/2\right)$,
  the thick constant line is $\alpha_{1/2(N_2-6)}=0$, and the thick dashed line is
  the line $\Gamma$ defined by Eq. (\ref{Gamma}). The gray area is the region in $\vartheta_{1}$-invariant set where bound entangled states are detected by the Breuer's map.
  The point where the dashed lines intersect is the point $\tilde{D}''$.}\label{6xN}
\end{figure*}

We illustrate our previous considerations by an example from $6\ot
N$ space (see Fig. \ref{6xN}). In this case the hyperplains
defined by (\ref{hyper}) become straight lines. One should notice
that the BE region (gray area in Fig. \ref{6xN}) detected by the
Breuer's map (\ref{B-op}) shrinks with the growth of N.
\section{Conclusion}
\label{Sec.IV}

In the present paper we have shown that in
$\mathbb{C}^{N_{1}}\ot\mathbb{C}^{N_{2}}$ Hilbert space of
rotationally invariant states with even $N_1 \geq 4$ and arbitrary
$N_2\geq N_1$ there always exist bound entangled states among PPT
states. This means simultaneously that partial transposition
provides only a necessary criterion for separability in these cases.
However, the problem is still unsolved for rotationally invariant
states acting on $\mathbb{C}^{N_{1}}\ot\mathbb{C}^{N_{2}}$ with odd
$N_{1} \geq 5$ and arbitrary $N_{2}\geq N_{1}$. Some preliminary
results suggest that bound entanglement do exist in such systems.
However, this issue requires further research.


We have described in more details the cases of $N_1=4$ and
$N_1=6$. The provided examples give the geometrical description of
the action of the map developed recently by Breuer
\cite{B-criterion} in the space of rotationally invariant states.
Moreover they reveal that the BE PPT region detected by Breuer's
map shrinks with the growth of asymmetry between the subsystems.
Moreover, the example of $4\ot N$ shows that in this case the
Breuer's map provides a necessary and sufficient criterion for
$SO(3)$, $\vartheta_1$-invariant states. This suggests that for
higher dimensional states of the form (\ref{thetainv}) the region
of entanglement might be unambiguously determined with the
criterion based on the $\Phi_1$ map.

Let us discuss entanglement detection of the rotationally
invariant states in the context of a recently proposed
separability criterion involving determinant of the partially
transposed density matrix \cite{det}. Namely, it says that a given
two-qubit state $\rho$ is separable if and only if
\begin{equation}\label{Conlusion1}
\det\rho^{PT}\geq 0,
\end{equation}
where $PT$ denotes standard partial transposition with respect to
an arbitrary subsystem. As shown in Ref. \cite{Schliemann1} and
then in Ref. \cite{3xN} partial transposition is also a sufficient
criterion for separability for rotationally invariant states
acting on $\mathbb{C}^{2}\ot\mathbb{C}^{N}$. Therefore one may
expect that the above criterion applies also to this class of
states. Using simple arguments it may be shown that this, indeed,
is the case for even $N$.

Such states as well as their partial time reversal can be written
in the form (\ref{alpha}) and their eigenvalues are proportional
to $\alpha_J$'s. For such systems $\alpha_{N/2}$ of a state after
partial time reversal (unitarily equivalent to partial
transposition) is positive. This means that if a given
rotationally invariant state is entangled its partial time
reversal must have $\alpha_{(N-2)/2}<0$, and therefore, in case of
even $N$, odd number of negative eigenvalues (eigenvalues of $2\ot
N$ rotationally invariant states with even $N$ have odd
degeneracy). of a density matrix after partial time reversal must
be negative.

\section*{Acknowledgments}
Authors are grateful to P. Horodecki for fruitful discussions and
commenting on the manuscript. Discussions with M. Czachor are
acknowledged. This work was supported by EU-IP Programme SCALA
(contract No. 015714) and Polish Ministry of Scientific Research
and Information Technology under the (solicited) grant no.
PBZ-MIN- 008/P03/2003.

\appendix

\section{}\label{AppA}
In this appendix we show that point $\tilde{D}''$ (\ref{de})
indeed belongs to the $\Gamma$ hyperplain (\ref{Gamma}). To show
this we rewrite Eq. (\ref{Gamma}) in the following form (here for
simplicity we use $j_{1(2)}$ instead of $(N_{1(2)}-1)/2$ whenever
it leads to shorter formulas)
\begin{equation}\label{App1}
(-1)^{N_2}\sum_{K=0}^{(N_1-2)/2}\sqrt{4K+1}\left\{\begin{array}{ccc}
  j_1 & j_2 & j_2-j_1 \\
  j_2 & j_1 & 2K\\
\end{array}\right\}\beta_{2K}+\frac{\sqrt{N_1 N_2}}{2N_2}=0.
\end{equation}
Now inserting coordinates of $\tilde{D}''$ given by Eq. (\ref{de})
we obtain:
\begin{equation}\label{App2}
\sum_{K=0}^{(N_{1}-2)/2}(4K+1)\left\{\begin{array}{ccc}
  j_1 & j_2 & j_{2}-j_{1} \\
  j_2 & j_1 & 2K\\
\end{array}\right\}
\left\{\begin{array}{ccc}
  j_1 & j_2 & j_{1}+j_{2} \\
  j_2 & j_1 & 2K \\
\end{array}\right\}+\frac{1}{2N_2}=0.
\end{equation}
We complete the sums on the left hand side with odd elements and
receive:
\begin{eqnarray}\label{App3}
&&\sum_{K=0}^{N_{1}-1}(2K+1)\left\{\begin{array}{ccc}
  j_1 & j_2 & j_{2}-j_{1} \\
  j_2 & j_1 & K\\
\end{array}\right\}
\left\{\begin{array}{ccc}
  j_1 & j_2 & j_{1}+j_{2} \\
  j_2 & j_1 & K \\
\end{array}\right\}
+\sum_{K=0}^{N_{1}-1}(-1)^{K}(2K+1)\left\{\begin{array}{ccc}
  j_1 & j_2 & j_{2}-j_{1} \\
  j_2 & j_1 & K\\
\end{array}\right\}
\left\{\begin{array}{ccc}
  j_1 & j_2 & j_{1}+j_{2} \\
  j_2 & j_1 & K \\
\end{array}\right\}\nonumber\\
&&=-1/N_2.
\end{eqnarray}
Both sums on the left-hand side of the above equation can be
calculated with the help of the following relations \cite{Edmonds}
\begin{equation}\label{App4}
\sum_{K}(2J+1)(2K+1)\left\{\begin{array}{ccc}
  a & b & J \\
  c & d & K \\
\end{array}\right\}\left\{\begin{array}{ccc}
  a & b & J' \\
  c & d & K \\
\end{array}\right\}=\delta_{JJ'}
\end{equation}
and
\begin{equation}\label{App5}
\sum_{K}(-1)^{K}(2K+1)\left\{\begin{array}{ccc}
  a & b & J \\
  c & d & K \\
\end{array}\right\}\left\{\begin{array}{ccc}
  a & c & J' \\
  b & d & K \\
\end{array}\right\}=(-1)^{J+J'}\left\{\begin{array}{ccc}
  a & b & J \\
  d & c & J' \\
\end{array}\right\}.
\end{equation}
The first sum in Eq. (\ref{App3}) equals 0 by the orthogonality
relation (\ref{App4}), and the second sum reduces to:
\begin{equation}
\sum_{K=0}^{N_{1}-1}(-1)^{K}(2K+1)\left\{\begin{array}{ccc}
  j_1 & j_2 & j_{2}-j_{1} \\
  j_2 & j_1 & K\\
\end{array}\right\}
\left\{\begin{array}{ccc}
  j_1 & j_2 & j_{1}+j_{2} \\
  j_2 & j_1 & K \\
\end{array}\right\}=(-1)^{2j_2}\left\{\begin{array}{ccc}
  j_1 & j_2 & j_{2}-j_{1} \\
  j_1 & j_2 & j_{1}+j_{2} \\
\end{array}\right\}=-\frac{1}{2j_2+1}=-\frac{1}{N_2},
\end{equation}
which is the right hand side of Eq. (\ref{App3}).$\blacksquare$

\end{document}